\pdfoutput=1

\documentclass[11pt,nofootinbib]{article}
\usepackage{amsmath,amssymb,mathtools}
\usepackage{graphicx}
\usepackage[colorlinks=true,linkcolor=blue,citecolor=blue,urlcolor=blue,linktocpage=true]{hyperref}
\usepackage{subfigure}
\usepackage{comment}
\usepackage{tensor}
\usepackage{cases}
\usepackage[utf8]{inputenc}
\usepackage{cite}
\usepackage[font=small,labelfont=bf]{caption}

\numberwithin{equation}{section}

\begin{document}

\providecommand{\abs}[1]{\lvert#1\rvert}
\providecommand{\bd}[1]{\boldsymbol{#1}}

\begin{titlepage}

\setcounter{page}{1} \baselineskip=15.5pt \thispagestyle{empty}

\begin{flushright}
\end{flushright}
\vfil

\bigskip
\begin{center}
 {\LARGE \textbf{Impact of Helical Electromagnetic Fields}}\\
 \medskip
 {\LARGE \textbf{on the Axion Window}}
\vskip 15pt
\end{center}

\vspace{0.5cm}
\begin{center}
{\Large
Takeshi Kobayashi$^{\star}$
and
Rajeev Kumar Jain$^{\dagger}$
}\end{center}

\vspace{0.3cm}

\begin{center}
\textit{$^{\star}$ Kobayashi-Maskawa Institute for the Origin of Particles and the Universe,\\ Nagoya University, Nagoya 464-8602, Japan}\\

\vskip 14pt 
\textit{$^{\dagger}$ Department of Physics, Indian Institute of Science, Bangalore 560012, India}\\
 
\vskip 14pt
E-mail:
 \texttt{\href{mailto:takeshi@kmi.nagoya-u.ac.jp}{takeshi@kmi.nagoya-u.ac.jp}},
 \texttt{\href{mailto:rkjain@iisc.ac.in}{rkjain@iisc.ac.in}}
\end{center} 



\vspace{1cm}

\noindent
Primordial electromagnetic fields can strongly affect the cosmic evolution of axions, and vice versa. We show that if helical electromagnetic fields are coherently produced in the early universe, their remnants source a field velocity to the coupled axions and enhance the relic abundance of axion dark matter. We discuss the implications for the QCD axion and axion-like particles that are coupled to the SM or hidden gauge groups. For a QCD axion coupled to hidden photons, we find that the conventional window for the axion decay constant $10^{8}\, \mathrm{GeV} \lesssim f \lesssim 10^{12}\, \mathrm{GeV}$ can be completely closed due to overproduction of axion dark matter by helical electromagnetic fields as little as $\alpha \, \Delta N_{\mathrm{eff}} \gtrsim 10^{-12}$, where $\alpha$ is the gauge coupling and $\Delta N_{\mathrm{eff}}$ is the effective extra relativistic degrees of freedom of the hidden photons.
\vfil

\end{titlepage}

\newpage
\tableofcontents

\section{Introduction}
\label{sec:intro}

Physics beyond the Standard Model (SM) can contain additional 
approximate global U(1) symmetries, that are spontaneously broken to
yield pseudo-Nambu--Goldstone bosons (PNGBs). 
Perhaps the most famous example is the QCD axion~\cite{Weinberg:1977ma,Wilczek:1977pj} arising from the
breaking of the Peccei--Quinn U(1) symmetry~\cite{Peccei:1977hh}
introduced for solving the strong $CP$ problem.
Low-energy effective field theories emerging
from string theory also suggest the presence of axion-like
particles~\cite{Svrcek:2006yi,Douglas:2006es,Arvanitaki:2009fg}.
Such PNGBs, which we collectively refer to as axions, are typically
light and couple very weakly to normal matter,
and thus provide a viable dark matter candidate. 

The QCD axion obtains its mass through the coupling $\theta G \tilde{G}$
to the gluons. 
However in general, axions may possess couplings 
$\theta F \tilde{F}$ with other gauge fields either within the SM or in
a hidden sector, Abelian or non-Abelian. 
Here, the product of the gauge field strength and its dual
corresponds to a dot product of the electric and magnetic fields, 
i.e., $F \tilde{F} \propto E \cdot B$.
This indicates that the dynamics of an axion can be modified in a
background of helical electromagnetic fields. 

Recently there is growing interest in the possibility that gauge fields
were excited in the early universe.
Cosmological production of the SM photon has long been a topic of study
with the aim of explaining the origin of the galactic and intergalactic
magnetic fields~\cite{Turner:1987bw, Ratra:1991bn, Subramanian:2015lua}. 
Nowadays, various kinds of gauge fields are
also studied
as a source of gravitational waves,
as dark radiation, vector dark matter,
in the context of anisotropic inflation,
and so on.
There is also a large literature on 
production mechanisms that spontaneously violate parity, 
in which case the resulting gauge particles can form helical
electromagnetic 
fields;
see e.g. \cite{Garretson:1992vt,Anber:2006xt,Durrer:2010mq,Byrnes:2011aa,Adshead:2016iae}.
Such a parity-odd gauge field background will, in turn, have a non-trivial effect
on axions.

In this paper we analyze the dynamics of the QCD axion and axion-like
fields in the presence of helical electromagnetic fields. 
If the helical fields are coherently excited
in the early universe, it gives an effective linear potential to the
axion via the coupling $\theta F \tilde{F}$, 
which forces the axion field to roll in one direction. 
We show that the $F \tilde{F}$ background has the net effect of increasing
the relic abundance of axion dark matter, and therefore can completely
transform the parameter window where axions are compatible with cosmology.
Since an $E \cdot B$ composed of the SM photons vanishes in the
reheating epoch as the conductivity of the universe rises
and the electric fields short out, in this paper
we will mainly focus on 
U(1) gauge fields in a hidden sector,
and show that hidden electromagnetic fields can significantly modify
the axion window.
However, the SM electromagnetic fields can also have strong effects
if the reheating temperature is sufficiently low.

Before proceeding with our analysis, 
we should remark that the possibility of helical fields
affecting the axion dynamics was first pointed out 
in~\cite{Campanelli:2005jw}.
However this work studies the effect in a universe with vanishing
electric fields and argues that the axion obtains a significant velocity
by helical magnetic fields alone,
even though in such a case there is no coherent $F\tilde{F}$ and so
there should be no source for the axion velocity.
We explicitly show that electric fields are necessary for affecting the
axion dynamics, in disagreement with~\cite{Campanelli:2005jw}. 
Similar conclusions on this point were reached
in~\cite{Long:2015cza,Dvornikov:2020hft}, 
which studied helical SM magnetic fields in a conducting cosmological
plasma (and hence with tiny electric fields)
and found the effect on the QCD axion to be negligible. 
Let us also mention that, 
besides electromagnetic fields, there can be other effects
that source a velocity to the axion field in the early universe
and modify the conventional axion window.
These include
axion potentials with multiple hierarchically separated periods~\cite{Choi:2015fiu,Kaplan:2015fuy},
shift-symmetric couplings to gravity~\cite{DeSimone:2016bok},
derivative couplings to other coherent scalar fields such as the
inflaton~\cite{Kobayashi:2019eyg,Kobayashi:2020ryx},
and an explicit breaking of the shift symmetry by higher-dimensional
operators~\cite{Kamionkowski:1992mf,Holman:1992us,Co:2019jts}.

This paper is organized as follows:
In Section~\ref{sec:dynamics} we give general discussions on the
axion dynamics in a background of helical electromagnetic fields.
In Section~\ref{sec:abundance} we explicitly compute the 
effect of the helical fields on the relic abundance of axion-like
particles and the QCD axion, and show how the parameter windows are
affected. 
We conclude with a discussion of directions for further research 
in Section~\ref{sec:conc}.
In the appendix we analyze the cosmological evolution of helical
electromagnetic fields.

\section{Axion Dynamics with Helical Electromagnetic Fields}
\label{sec:dynamics}

We consider an axion coupled to a U(1) gauge field, 
\begin{equation}
 \frac{\mathcal{L}}{\sqrt{-g}} = 
-\frac{1}{2} f^2g^{\mu \nu} \partial_\mu \theta \partial_\nu \theta 
- m^2 f^2 \left( 1 - \cos \theta  \right)
+\frac{\alpha }{8 \pi }\theta F_{\mu \nu} \tilde{F}^{\mu \nu}.
\label{eq:action}
\end{equation}
Here the axion is written as a dimensionless angle~$\theta$,
in terms of which the distance between the adjacent vacua is $\Delta
\theta = 2 \pi$. 
The mass~$m$ may or may not depend on the cosmic temperature,
and $f$ is the axion decay constant.
The U(1) gauge field can be either the SM photon or a hidden photon, 
$\alpha$ is a dimensionless gauge coupling and can have either sign,
and the dual field strength is 
\begin{equation}
 \tilde{F}^{\mu \nu} = \frac{1}{2} \eta^{\mu \nu \rho
\sigma } F_{\rho \sigma}.
\label{eq:Ftilde}
\end{equation}
Here 
$\eta^{\mu \nu \rho \sigma} = \epsilon^{\mu \nu \rho \sigma} /\sqrt{-g}$
is a totally antisymmetric pseudotensor 
with the Levi--Civita symbol normalized as
$\abs{\epsilon^{0123}} = 1$.

Throughout this paper we consider cases where the global U(1) is already
broken by the end of inflation, and continues to be broken in the
post-inflationary epoch. Thus we impose
\begin{equation}
 f >  \frac{H_{\mathrm{inf}}}{2 \pi }, T_{\mathrm{max}},
\label{eq:maru-1}
\end{equation}
where $H_{\mathrm{inf}} $ is the Hubble rate during inflation,
$T_{\mathrm{max}}$ is the highest temperature achieved in the
post-inflationary universe, and for 
simplicity we identified the symmetry breaking scale with the decay
constant. 
The inflationary expansion
sets the axion field to be spatially
homogeneous throughout the observable universe, 
and gives rise to the vacuum misalignment
scenario~\cite{Preskill:1982cy,Abbott:1982af,Dine:1982ah} for axion dark
matter. 

However, the basic picture of the vacuum misalignment becomes modified 
in the presence of a coherent $F\tilde{F}$ background,
because the $\theta F \tilde{F}$ coupling sources an
effective linear potential for the axion.
This is also seen in the axion's equation of motion. 
Fixing the metric to a flat FRW, 
\begin{equation}
 ds^2 =  -dt^2 + a(t)^2 d\bd{x}^2,
\end{equation}
the equation of motion for a homogeneous axion field reads
\begin{equation}
 0 = \ddot{\theta} + 3 H \dot{\theta} + m^2 \sin \theta - \frac{\alpha
  }{8 \pi } \frac{F\tilde{F}}{f^2}.
\label{eq:EoM}
\end{equation}
Here, an overdot denotes a $t$-derivative, and $H = \dot{a} / a$.
Before studying this equation in detail, we first discuss the physical
meaning of $F\tilde{F}$.

\subsection{Helical Electromagnetic Fields in the SM and Hidden Sector}

The term $F\tilde{F}$ can be written as a dot product
of electric and magnetic fields,
\begin{equation}
 F_{\mu \nu} \tilde{F}^{\mu \nu} = -4 E_\mu B^\mu,
\end{equation}
where 
\begin{equation}
 E^\mu = u_\nu F^{\mu \nu}, \quad
 B^\mu = \frac{1}{2} \eta^{\mu \nu \rho \sigma} u_\sigma F_{\nu \rho},
\label{eq:EandB}
\end{equation}
are the fields measured by a comoving observer with
4-velocity $u^\mu$ ($u_\mu u^\mu = -1$, $u^i = 0$; we use Latin letters
to denote spatial indices). 
Even for a hidden U(1), we refer to the quantities in
(\ref{eq:EandB}) as the `electric' and `magnetic' fields.
As can be understood by noting that $F \tilde{F}$ is parity-odd, a
nonvanishing $E \cdot B$ implies an asymmetry between the two circular
polarization states, and hence such electromagnetic fields are referred
to as `helical' fields.

The energy density in the gauge field is a sum of
the electromagnetic fields squared,\footnote{This can be 
checked by varying the gauge kinetic term $-FF/4$ 
with respect to the metric to obtain the energy-momentum tensor. 
Note that the interaction term $\theta F
\tilde{F}$ does not contribute to the energy-momentum tensor.}
$ \rho_A = ( E_\mu E^\mu + B_\mu B^\mu ) / 2$.
Then, noting that 
$(E_\mu \pm B_\mu) (E^\mu \pm B^\mu) \geq 0$, one finds that the
magnitude of the product $E \cdot B$ is bounded by the energy density,
\begin{equation}
\abs{E_\mu B^\mu}
\leq \rho_A\, .
\label{eq:12.7}
\end{equation}

Even if an electric field composed of the
SM photons (or hypercharge gauge bosons) is produced in the
early universe, it gets shorted out during the reheating epoch 
as the conductivity of the universe rises,
after which $F\tilde{F}$ also vanishes.
(For a discussion on the evolution of the
conductivity, see e.g. \cite{Turner:1987bw}.)
However, SM electric fields may still survive
until rather late times, 
if the reheating temperature is low, 
and also if the decay of the inflaton happens abruptly as in some models of
preheating~\cite{Kofman:1997yn} instead of gradually through a perturbative decay.
On the other hand, primordial electric fields composed of hidden photons
that are decoupled from the SM
remain intact during reheating, 
unless particles charged under the hidden U(1) are also produced.
In the following we will mainly consider helical
electromagnetic fields composed of hidden photons,
however the analyses will also apply to the SM photon until the time
when the electric field disappears.

If the U(1) gauge field is a hidden photon, then it behaves as extra
radiation and contributes to the 
effective extra relativistic degrees of freedom of the universe via
\begin{equation}
 \Delta N_A = \frac{8}{7} \left( \frac{11}{4} \right)^{4/3}
 \frac{\rho_A}{\rho_\gamma },
\end{equation}
with $\rho_\gamma$ being the energy density in the SM photon.
With this expression in mind, 
we can parameterize the 
amplitude of the dot product of the electric and magnetic fields as
\begin{equation}
 \Delta N_{E \cdot B} \equiv \frac{8}{7} \left( \frac{11}{4} \right)^{4/3}
\frac{\abs{E_\mu B^\mu }}{\rho_\gamma }
\leq \Delta N_A 
\lesssim 10^{-1},
\label{eq:NEB}
\end{equation}
where the first inequality arises from (\ref{eq:12.7}) and is saturated
for maximally helical fields.
The second inequality shows the current constraint on extra radiation
from CMB measurements~\cite{Aghanim:2018eyx}.
After the electron-positron annihilation, 
the SM photon energy density redshifts as $\rho_\gamma \propto a^{-4}$.
Hence if $E \cdot B$ also redshifts in a radiation-like manner 
of $ E \cdot B \propto a^{-4}$, then 
$\Delta N_{E \cdot B}$ is time-independent.\footnote{If the dot product
continues to redshift as $E_\mu B^\mu \propto a^{-4}$ until today, 
its present-day amplitude in Gauss 
(although these are electromagnetic fields in a hidden sector) is
$\abs{E_\mu B^\mu}_0 \sim (10^{-6} \, \mathrm{G})^2 \Delta N_{E \cdot B} $
in Heaviside--Lorentz units.}
However $E \cdot B$ can also exhibit other redshifting behaviors,
depending on how the helical fields were originally produced.

In the following analyses we assume the helical electromagnetic fields
to have been coherently produced in the early universe, and discuss
their consequence for axions. 
A toy example of a gauge field theory that produces
helical fields is discussed in
Appendix~\ref{app:evolution}, where we also analyze the redshifting
behaviors of the helical fields after being produced.

\subsection{Induced Axion Velocity}

Upon solving the equation of motion~(\ref{eq:EoM}), 
let us for the moment ignore the axion potential, namely, we set $m = 0$. 
We further suppose that the background universe has a constant
equation of state~$w$, and that 
$F\tilde{F}$ is homogeneous and redshifts with some power of the scale
factor, 
\begin{equation}
H \propto a ^{-\frac{3 (w+1)}{2}},
\quad
 F \tilde{F} \propto a^{-n}.
\label{eq:powers}
\end{equation}
Then the equation of motion (\ref{eq:EoM}) can be solved to yield the
axion velocity as
\begin{equation}
 \dot{\theta} =
\left\{ - n + \frac{3 (w+3)}{2} \right\}^{-1}
\frac{\alpha }{8 \pi } \frac{F \tilde{F}}{f^2 H}
+ K a^{-3},
\label{eq:a2p}
\end{equation}
where we have assumed $n \neq 3 (w+3)/2$, and $K$ is an arbitrary
constant. 
Without the source term, i.e. $ F\tilde{F} = 0$, 
any initial velocity quickly decays away as
$\dot{\theta} \propto a^{-3}$
as for any massless scalar in an expanding universe.
However in the presence of an $F\tilde{F}$ background, the axion picks up
a contribution $ \dot{\theta} \propto \alpha F \tilde{F} / f^2 H$.
This redshifts slower than $ a^{-3}$ and eventually dominates the axion
velocity if $n < 3 (w+3) / 2$, which is satisfied, e.g., with $n = 4$ in
a decelerating universe.\footnote{In the main part of the work 
\cite{Campanelli:2005jw}, electric fields are 
considered to be zero and so is $F\tilde{F}$,
and hence they make no mention of the specific solution 
$\abs{\dot{\theta}} \sim \abs{\alpha F \tilde{F} / 8 \pi f^2 H}$.
The problem with their analysis is that they  
fix the integration constant by hand such that it depends on the 
magnetic helicity~$\mathcal{H}_B$ as 
$K \propto \alpha \mathcal{H}_B / f^2 $.
In this way they arrive at the incorrect conclusion that an axion
velocity is induced by helical magnetic fields alone.
They further argue that the relic abundance of the QCD axion depends
linearly on the magnetic helicity, however we show in
Section~\ref{sec:QCDaxion} that it actually depends on the amplitude of
$E \cdot B$ with fractional powers.}

The assumption of a negligible axion potential is
justified if, in the equation of motion~(\ref{eq:EoM}),
the tilt of the $F\tilde{F}$-induced linear potential is
larger than that of the axion potential.
This condition is written under $\abs{\sin \theta} \sim 1 $ as
\begin{equation}
 \left| \frac{\alpha }{8 \pi } \frac{F\tilde{F}}{f^2}  \right| >  m^2.
\label{eq:m2}
\end{equation}
The axion potential can be neglected also if the induced
kinetic energy of the axion 
$ \rho_{\theta \, \mathrm{kin}} =  f^2 \dot{\theta}^2 /2 $
is larger than the height of the periodic axion potential~$2 m^2 f^2$,
i.e., 
\begin{equation}
 \left| \frac{\alpha }{8 \pi } \frac{F\tilde{F}}{f^2}  \right| > m H ,
\label{eq:mH}
\end{equation}
where for simplicity we have assumed $|n - 3 (w+3)/2| \sim 1$ and $K = 0$
in (\ref{eq:a2p}) for the axion velocity, and dropped order-unity factors.
This condition implies that the axion 
is indeed able to go over the maxima of its periodic potential.

One may wonder what happens when only one of 
(\ref{eq:m2}) or (\ref{eq:mH}) is satisfied.
In the case of
$ m H < \abs{\alpha  F \tilde{F} / 8 \pi f^2} < m^2 $ 
(which necessarily entails $ H < m$),
the time it takes for the axion to move over the period of its
potential 
$\Delta t = 2 \pi / \abs{\dot{\theta}}$ with velocity
$\abs{\dot{\theta}} \sim \abs{\alpha F \tilde{F} / 8 \pi f^2 H}$,
is shorter than the oscillation period $ 2 \pi / m$ along the axion
potential. 
This suggests that by averaging over a time interval shorter than 
$ 2 \pi / m$, the axion potential in the equation of motion vanishes, 
i.e. $\langle m^2 \sin \theta \rangle \approx 0$,
and thus it can be neglected.
On the other hand if
$ m^2 < \abs{\alpha  F \tilde{F} / 8 \pi f^2} < m H $,
the axion moves less than~$2 \pi$ in a Hubble time;
in this case the axion moves with velocity 
$\abs{\dot{\theta}} \sim \abs{\alpha F \tilde{F} / 8 \pi f^2 H}$
but is trapped in a potential well over cosmological time scales.

Thus in summary, 
a coherent $F\tilde{F}$ background sources an axion velocity of 
\begin{equation}
  \abs{\dot{\theta}} \sim  
\left| \frac{\alpha }{8 \pi } \frac{F \tilde{F}}{f^2 H}\right|,
\label{eq:axv}
\end{equation}
given that its amplitude is as large as
\begin{equation}
 \left| \frac{\alpha }{8 \pi } \frac{F\tilde{F}}{f^2}  \right| >
\mathrm{min.}\{ m^2, mH  \}.
\label{eq:cond-min}
\end{equation}

\subsection{Backreaction from the Axion}

The axion can backreact to the gauge field, 
as a rolling axion itself induces excitation of the coupled gauge field. 
The effect becomes significant when the
factor in front of $F \tilde{F}$ in the action~(\ref{eq:action})
varies by a factor of order unity or larger within a Hubble
time~\cite{Garretson:1992vt,Anber:2006xt,Durrer:2010mq,Byrnes:2011aa}.
This condition can be written, using (\ref{eq:axv}), as 
\begin{equation}
\left| \frac{\alpha }{8 \pi } \frac{\dot{\theta} }{H}  \right| \sim
 \left(\frac{\alpha }{8 \pi} \right)^2
  \frac{ \abs{F \tilde{F}} }{f^2 H^2} 
> 1.
\label{eq:17.2}
\end{equation}
This is equivalent to saying that the induced
kinetic energy of the axion is larger than~$\abs{F\tilde{F}}$. 
A coherent $F\tilde{F}$ as large as~(\ref{eq:17.2}) moves the axion
rapidly in one direction, which in turn is expected to produce 
gauge bosons with momenta typically of order the Hubble scale at that time.
This process should slow down the axion velocity,
and also implies the cascading of the power of $F\tilde{F}$ 
towards higher momenta. 
A detailed study of the axion electrodynamics in such a regime will
likely require lattice studies and is beyond the scope of this work, 
however it would be very interesting to investigate the possibility of 
the axion-induced UV cascade of helical electromagnetic
fields.\footnote{If $\abs{(\alpha/8\pi) (\dot{\theta} / H)}$ is
sufficiently larger than unity during inflation, it can also generate
large non-Gaussianities in the curvature perturbation and/or violate
perturbativity, see e.g. \cite{Ferreira:2014zia,Ferreira:2015omg}. 
We also note that, while here we discussed the backreaction from the
rolling axion, gauge field excitations may also happen during the
oscillatory phase~\cite{Agrawal:2017eqm,Kitajima:2017peg}.}

\subsection{Onset of Axion Oscillation}

In the conventional vacuum misalignment scenario
without an $F\tilde{F}$ background, 
the axion field stays frozen at some initial value
while $H > m$, then begins to oscillate about a potential minimum when
$H \sim m $.
Henceforth we use the subscript ``$m$'' to denote quantities at the time 
the Hubble rate becomes equal to the axion mass
(note that the mass may also vary in time),
\begin{equation}
 H_m = m_m.
\label{eq:a_m}
\end{equation}
The energy density of the axion at the onset of the oscillation is 
\begin{equation}
 \rho_{\theta m} = b \, (m^2  f^2 \theta^2)_m ,
\end{equation}
where the cosine potential has been expanded around zero
assuming $\abs{\theta_m} \lesssim 1$, 
and $b$ is a factor of order unity. 
From then on, the axion oscillates and its particle number is conserved,
so the physical number density $n_\theta = \rho_\theta / m $
redshifts as $ a^{-3}$.
The relic abundance today is thus obtained as
\begin{equation}
 \rho_{\theta 0} = m_0 n_{\theta 0} = b\,  
m_0 m_m f^2 \theta_m^2 \left( \frac{a_m}{a_0} \right)^3 ,
\label{eq:rho_conv}
\end{equation}
where the subscript ``$0$'' is used to denote quantities in the present
universe.  

Now, in the presence of an $F\tilde{F}$ background, 
given that both $H/m$ and $\abs{F\tilde{F} / m H}$ monotonically
decrease in time, 
there are two possible scenarios.
The first is the case where the condition~(\ref{eq:cond-min}) 
is never satisfied after the time of $ H = m $.
It may have been satisfied while $H > m$, however this 
merely moves the axion field prior to the onset of the oscillation.
Hence the effect of $F\tilde{F}$ can be absorbed by a shift
in the value of~$\theta_m$, and the axion's relic abundance is given
by~(\ref{eq:rho_conv}).

The story is drastically modified if the
condition~(\ref{eq:cond-min}) continues to hold after the
time of $ H = m$, namely, if
\begin{equation}
 \left| \frac{\alpha }{8 \pi } \frac{(F\tilde{F})_m}{f^2}  \right|
>  m^2_m.
\label{eq:iv}
\end{equation}
In this case the axion continues to move in one direction 
with the velocity~(\ref{eq:axv}) even at times when $H < m$, until 
the condition (\ref{eq:cond-min}) is saturated
when $\abs{\alpha F \tilde{F} / 8 \pi f^2}$ 
becomes equal to $mH$; 
we denote quantities then by the subscript~``tr'', namely,
\begin{equation}
 \left| \frac{\alpha }{8 \pi } \frac{(F\tilde{F})_{\mathrm{tr}}}{f^2}  \right|
=  (m H)_{\mathrm{tr}}\, .
\label{eq:trap}
\end{equation}
At this time the axion's kinetic and potential energies become
comparable, and hence the axion gets trapped in its potential well and
begins to oscillate about a minimum.
Supposing that upon trapping the axion field is displaced from the
nearest potential minimum by
$\abs{\theta_{\mathrm{tr}} - \theta_{\mathrm{min}}} \sim 1$, 
the axion's energy density is written as
\begin{equation}
 \rho_{\theta \, {\mathrm{tr}}} = c\,  m_{\mathrm{tr}}^2 f^2,
\end{equation}
with $c$ being a factor of order unity.
Hereafter we can follow the same steps as for the conventional
misalignment scenario, and obtain the relic abundance as
\begin{equation}
 \rho_{\theta 0} = c\,  m_0 m_{\mathrm{tr}} f^2 \left(
   \frac{a_{\mathrm{tr}}}{a_0} \right)^3 .
\label{eq:v}
\end{equation}
Comparing this result with (\ref{eq:rho_conv}), one sees that 
the net effect of an $F\tilde{F}$ background satisfying
(\ref{eq:iv}) is to enhance the axion relic abundance by 
delaying the onset of the oscillation 
from when $H \sim m$ to the time of trapping described by~(\ref{eq:trap}).

We should remark that the approximation (\ref{eq:v}) with $ c \sim
1$ can break down if the coefficient 
$\abs{n - 3 (w+3)/2}^{-1}$ in (\ref{eq:a2p}) that we have dropped is
much larger (smaller) than unity; in such cases the time of the trapping 
will deviate from~(\ref{eq:trap}),
leading to an enhancement (suppression) of the final abundance.
In addition, if
$ \abs{\theta_{\mathrm{tr}} - \theta_{\mathrm{min}}} \approx \pi$, 
then the relic abundance will be enhanced by anharmonic
effects~\cite{Turner:1985si,Bae:2008ue} (the same is true for the
expression~(\ref{eq:rho_conv}) for the conventional scenario).

\subsection{Numerical Example}

\begin{figure}[t]
\begin{center}
 \begin{minipage}{.46\linewidth}
  \begin{center}
 \includegraphics[width=\linewidth]{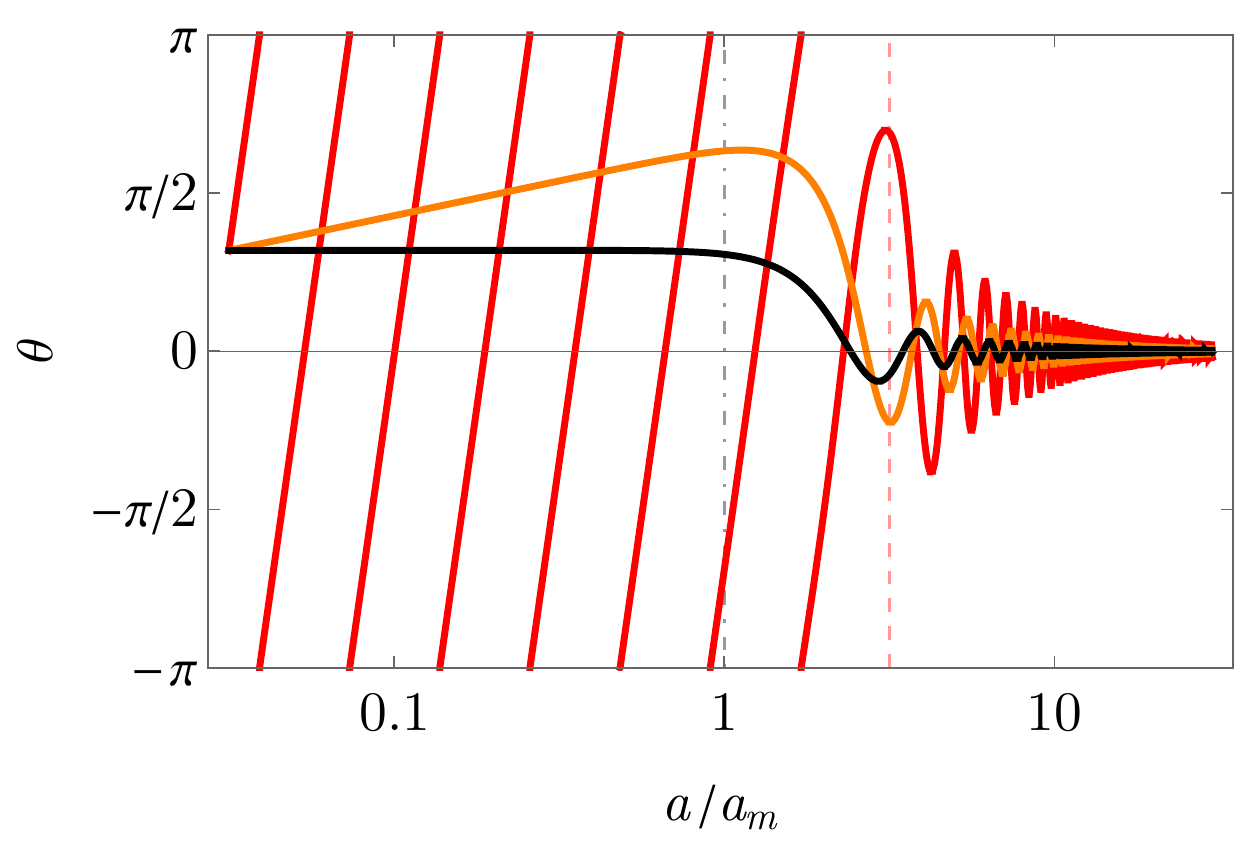}
  \end{center}
 \end{minipage} 
 \begin{minipage}{0.01\linewidth} 
  \begin{center}
  \end{center}
 \end{minipage} 
 \begin{minipage}{.46\linewidth}
  \begin{center}
 \includegraphics[width=\linewidth]{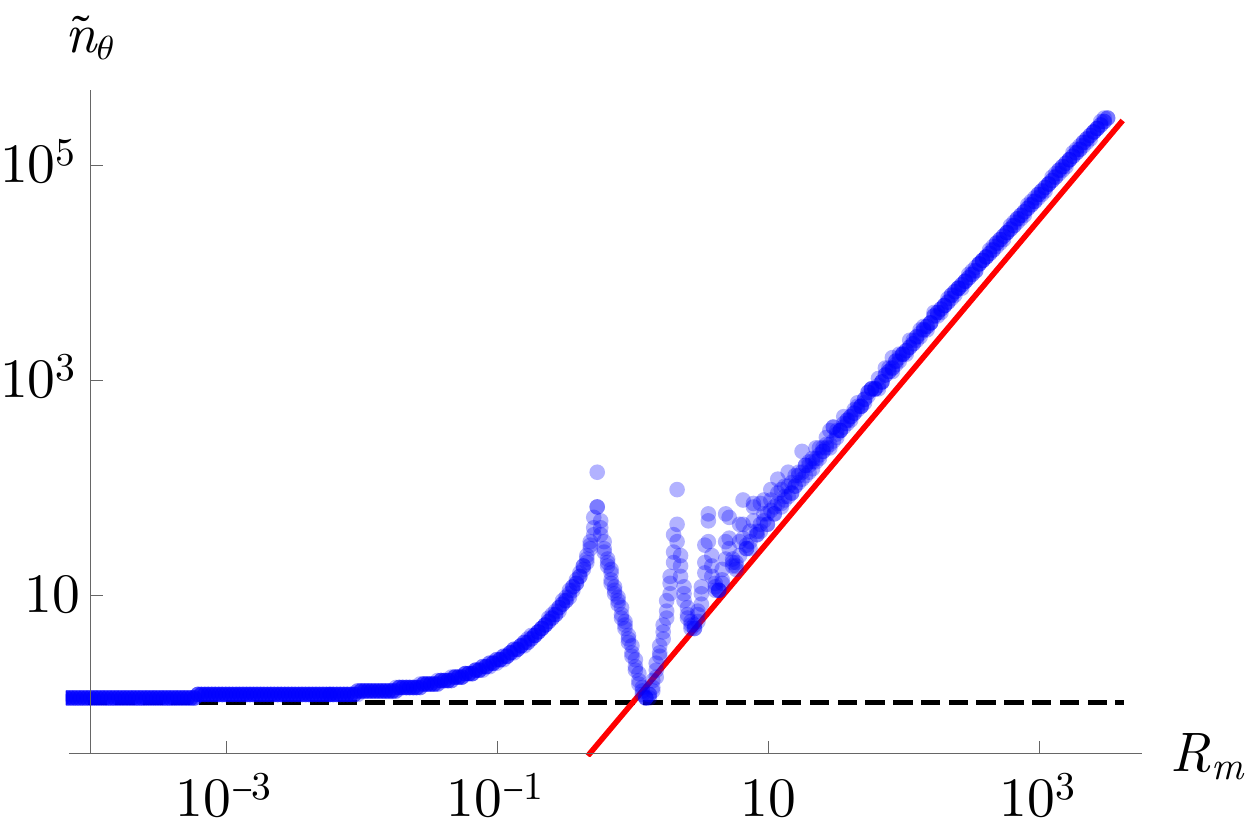}
  \end{center}
 \end{minipage} 
 \caption{Left: Time evolution of the axion angle as a function of the
 scale factor, in log-linear scale. 
 The upper and lower edges are identified. 
 The amplitude of $F\tilde{F}$ is varied, in terms
 of the normalized quantity~(\ref{eq:R_m}), as $R_m \ll 0.1$ (black
 curve), $R_m = 0.3$ (orange), and $R_m = 10$ (red). The dot-dashed
 vertical line shows when $a = a_m$, and the red dashed vertical line
 is $a = a_{\mathrm{tr}}$ for $R_m = 10$. 
 When $F\tilde{F}$ is sufficiently large, the axion is forced to move
 through multiple periods, and the onset of the axion oscillation is
 delayed. 
 Right: Normalized comoving number density of the axion in the asymptotic future,
 as a function of $R_m$. 
 The blue dots show the numerical results, while
 the black dashed and red solid lines show analytic estimates.
 See the text for details.}
 \label{fig:num-theta}
\end{center}
\end{figure}

We verified the axion dynamics discussed above by numerically
solving the equation of motion~(\ref{eq:EoM}), 
whose results are shown in Figure~\ref{fig:num-theta}.
Here we have set the background Hubble parameter and helical fields to 
redshift following the form of (\ref{eq:powers}),
with powers $w = 1/3$ and $n = 4$.
We took the axion mass to be a constant,
and started the computation from the time when $H = 10^3 m$,
with an initial condition $\theta_{\mathrm{ini}} = 1$ 
and velocity~(\ref{eq:a2p}) with $K = 0$. 
The time evolution of the axion field is shown in the left panel as
a function of $a/a_m$, in log-linear scale.
Here we have limited the displayed range to 
$-\pi \leq \theta \leq \pi$ by identifying the points $\theta = \pm \pi$, 
so that the minimum around which the axion eventually oscillates is
$\theta = 0$. 
Shown are three cases in which the $F\tilde{F}$ amplitude, parameterized as
\begin{equation}
 R_m = \left| \frac{\alpha }{8 \pi }
\frac{ (F \tilde{F})_m }{m^2 f^2}  \right|,
\label{eq:R_m}
\end{equation}
is chosen as 
$R_m \ll 0.1$ (black curve), $R_m = 0.3$ (orange), and $R_m = 10$ (red). 
This parameterization is introduced such that
the condition (\ref{eq:iv}) corresponds to $R_m > 1$.
The results displayed in this subsection are independent of the exact
values of $\alpha F\tilde{F}$, $m$, $f$, etc.
However we note that we have chosen the sign as $\alpha F\tilde{F} >
0$, which forces the axion to move towards the positive direction.
For $R_m \ll 0.1$, it is seen that the effect of $F\tilde{F}$ 
is negligible and the axion dynamics is the same as in the
conventional misalignment scenario,
in which the axion begins to oscillate at around $a
= a_m$, indicated by the dot-dashed vertical line. 
For $R_m = 0.3$, the axion also begins oscillating at $a \sim a_m$, 
however until the oscillation the $F\tilde{F}$ background forces the
axion to move away from its initial position.
The case of $R_m = 10$ satisfies (\ref{eq:iv}), and the axion moves
through multiple periods until the time $a = a_{\mathrm{tr}}$, which
is indicated by the red dashed vertical line. 
Further increasing $R_m$ yields an even larger separation between 
$a_m$ and $a_{\mathrm{tr}}$.

We also carried out the computations for a wide range of values for $R_m$; 
in the right panel we plot the comoving axion number density 
($\propto n_\theta a^3$) evaluated after the axion has begun its
oscillation and the number become conserved,
as a function of $R_m$ 
(the sign is chosen as $\alpha F\tilde{F} > 0$). 
The numerical results are shown as the blue dots, 
and the comoving number density in the $y$-axis is normalized as 
\begin{equation}
 \tilde{n}_\theta = \frac{n_\theta }{m f^2}
\left( \frac{a}{a_m} \right)^3.
\end{equation}
One can also analytically calculate this quantity
using (\ref{eq:rho_conv}) and (\ref{eq:v})
with $w = 1/3$, $n = 4$,
and by setting $\theta_{m} = b = c = 1$, as
\begin{numcases}{\tilde{n}_\theta  =  }
1
       & for $R_m < 1$, \label{222} \\
R_m^{\frac{6}{2 n-3 (w+1)}} = R_m^{3/2}
       & for $R_m \geq 1$. \label{333}
\end{numcases}
These are shown in the plot as the black dashed and
red solid lines, respectively.
One sees that the analytic expressions with $b = 1 $ and $ c = 1$ 
are consistent at the order-of-magnitude level with the numerical results 
in the asymptotic regions $R_m \ll 1$ and $R_m \gg 1$.

The oscillatory behavior of $\tilde{n}_\theta$, which is most prominent
around $R_m \sim 1$, is due to the $F\tilde{F}$ background shifting the
misalignment angle with a periodicity of~$2 \pi$;
such an effect is not captured in the analytic estimate
where we have fixed the misalignment to unity for simplicity.
The oscillation peaks of $\tilde{n}_\theta$ correspond to 
the axion being placed near the hilltop  $\theta =  \pm \pi$ at the onset
of the oscillation, which gives anharmonic enhancements to the relic
abundance.
We remark that, since we have carried out the computations
only for a finite number of values of~$R_m$, 
the plot does not fully uncover the shape of the anharmonic peaks.
Note also that in the extremely anharmonic region, axionic domain
walls are expected to form, and thus one will have to include spatial
inhomogeneities into the analyses.
We should also mention that, as one increases $R_m$ much beyond unity, 
the backreaction from the axion will become non-negligible at $R_m >
8\pi / \abs{\alpha}$, as can be seen from the criteria~(\ref{eq:17.2}).

\section{Axion Relic Abundance}
\label{sec:abundance}

\subsection{Constant-Mass Axion}
\label{sec:alp}

As the simplest example, we begin by considering an axion whose mass~$m$
is a constant parameter.
We assume this axion to be coupled to hidden photons that make up
helical electromagnetic fields.
Moreover, to make our discussion concrete, we assume the helical fields to
redshift in a radiation-like manner, i.e.,
\begin{equation}
 F \tilde{F} \propto a^{-4}.
\label{eq:FtF-4}
\end{equation}
We also suppose the universe at the time of $ H = m$
to be dominated by radiation,
and the entropy of the universe to be conserved thereafter,
so that the entropy density redshifts as $s \propto a^{-3}$. 
These assumptions allow us to write
$(F\tilde{F})_m = (F \tilde{F})_0 (s_m / s_0)^{4/3}$.
Further noting that the Hubble rate and the entropy density during
radiation domination are expressed in terms of the cosmic temperature as
\begin{equation}
 3 M_{\mathrm{Pl}}^2 H^2 \simeq \rho_{\mathrm{rad}} 
= \frac{\pi^2 }{30 }g_* (T)\,  T^4,
\quad
s = \frac{2 \pi^2}{45}g_{*s} (T)\,  T^3,
\label{eq:RD}
\end{equation}
the entropy density at $H=m$ can be written as
\begin{equation}
 s_m = \left( \frac{128 \pi^2}{45} \right)^{1/4}
\frac{g_{*s} (T_m)}{g_*(T_m)^{3/4}}
(m M_{\mathrm{Pl}} )^{3/2}.
\end{equation}
Thus the condition (\ref{eq:iv}) for the helical fields to delay the
onset of the axion oscillation 
is translated into a lower bound on the field values today,  
\begin{equation}
 \left| \frac{\alpha }{8 \pi } (F\tilde{F})_0  \right|
> \left(\frac{45}{128 \pi^2}\right)^{1/3}
\frac{g_*(T_m)}{g_{*s}(T_m)^{4/3}}
\frac{f^2 s_0^{4/3}}{M_{\mathrm{Pl}}^2}.
\label{eq:I-iv-ex}
\end{equation}
Under this condition $a_{\mathrm{tr}}  > a_m$,
and hence one can also solve for $s_{\mathrm{tr}}$ by combining 
(\ref{eq:trap}), that is,
\begin{equation}
m  H_{\mathrm{tr}} = 
 \left| \frac{\alpha }{8 \pi } \frac{(F\tilde{F})_{\mathrm{tr}}}{f^2}  \right|
=  
 \left| \frac{\alpha }{8 \pi } \frac{(F\tilde{F})_{0}}{f^2}  \right|
\left( \frac{s_{\mathrm{tr}}}{s_0} \right)^{4/3},
\label{eq:4.9}
\end{equation}
with (\ref{eq:RD}) which can be used to rewrite $H_{\mathrm{tr}}$
in terms of $s_{\mathrm{tr}}$.
Then substituting $ (a_{\mathrm{tr}} / a_0)^3 = s_0 / s_{\mathrm{tr}}$ 
into (\ref{eq:v}) gives
\begin{equation}
 \rho_{\theta 0}  = 
c \left( \frac{128 \pi^2}{45} \right)^{1/4}
\frac{g_{*s} (T_{\mathrm{tr}})}{g_{*} (T_{\mathrm{tr}})^{3/4}}
\left| \frac{ \alpha}{8 \pi } (F \tilde{F})_0  \right|^{3/2}
\frac{ m^{1/2} M_{\mathrm{Pl}}^{3/2}}{f s_0 },
\label{eq:I-ii-ex}
\end{equation}
where the power $3/2$ of the $\abs{\alpha F \tilde{F}}$ term corresponds
to that of $R_m$ in the expression~(\ref{333}). 
We can further rewrite $\abs{F \tilde{F}}_0$ 
in terms of $\Delta N_{E \cdot B}$ defined in (\ref{eq:NEB}),
by noting that $\Delta N_{E \cdot B}$ becomes constant 
after the electron-positron annihilation
due to the assumption of $ F\tilde{F} \propto a^{-4}$.
Then, plugging in numbers for the physical constants and
cosmological parameters,
and also setting $ c \sim  1$ and $g_{*s} / g_*^{3/4} \sim 1$,
the axion abundance (\ref{eq:I-ii-ex}) and the condition
(\ref{eq:I-iv-ex}) are written as
\begin{equation}\label{eq:I-ii}
\begin{split}
 \Omega_\theta h^2 \sim
10^{-1}
\left( \frac{\abs{\alpha} \Delta N_{E \cdot B}}{10^{-2}} \right)^{3/2}
\left(\frac{f}{10^{17}\, \mathrm{GeV}}\right)^{-1}
&\left( \frac{m}{10^{-22}\, \mathrm{eV}} \right)^{1/2}
\\
& \mathrm{for} \, \, \, 
\abs{\alpha} \Delta N_{E \cdot B} \gtrsim 10^{-2} 
\left(\frac{f}{10^{17}\, \mathrm{GeV}}\right)^2.
\end{split}
\end{equation}
Here the abundance is expressed in units of the present value of the
critical density, 
and $h$ is the dimensionless Hubble parameter.
Here and below $\Delta N_{E \cdot B}$ is the value after the
electron-positron annihilation, and note that the
axion abundance depends on the helical fields through the
combination~$\abs{\alpha} \Delta N_{E \cdot B}$.

On the other hand, if the amplitude of the helical fields is small
enough such that (\ref{eq:iv}) is not satisfied,
then the conventional vacuum misalignment scenario is recovered.
The abundance is computed as (\ref{eq:rho_conv}),
giving the familiar result~\cite{Hui:2016ltb},
\begin{equation}
 \Omega_\theta h^2 \sim
10^{-1} \, \theta_m^2
\left( \frac{f}{10^{17}\, \mathrm{GeV}} \right)^2
\left( \frac{m}{10^{-22}\, \mathrm{eV}} \right)^{1/2}
 \quad \mathrm{for} \, \, \,
\abs{\alpha} \Delta N_{E \cdot B} \lesssim 10^{-2} 
\left(\frac{f}{10^{17}\, \mathrm{GeV}}\right)^2.
\label{eq:alp-vm}
\end{equation}
Note that if $\abs{\alpha} \Delta N_{E \cdot B}$ is non-zero, 
the misalignment angle~$\theta_m$ when the axion
begins to oscillate can be different from the value at earlier times,
say, at the end of inflation.

One can collectively write the axion abundance in the two regimes 
(\ref{eq:I-ii}) and (\ref{eq:alp-vm}) as: 
\begin{equation}
 \Omega_\theta h^2 \sim
10^{-1} \left(\frac{m}{10^{-22}\, \mathrm{eV}}\right)^{1/2}
\left(\frac{f}{10^{17}\, \mathrm{GeV}}\right)^{-1}
\left[
\mathrm{max.}
\left\{
\left( \frac{\abs{\alpha } \Delta N_{E \cdot B}}{10^{-2}} \right), 
\left(\frac{f}{10^{17}\, \mathrm{GeV}}\right)^{2}
\right\}
\right]^{3/2},
\label{eq:I-ab}
\end{equation}
where we have ignored the possibility of a fine-tuned initial
angle and thus set $\theta_m^2 \sim 1$ in (\ref{eq:alp-vm}).
This expression clearly shows that helical electromagnetic fields, when
large enough, have the effect of increasing the axion abundance.

Let us also assess the backreaction from the axion to the gauge field. 
With $F\tilde{F} \propto a^{-4}$, the
ratio~$F\tilde{F} / H^2$ stays more or less constant during radiation
domination, hence we evaluate the backreaction condition~(\ref{eq:17.2})
at the time of the trapping~(\ref{eq:trap}).
(Note that this condition can be satisfied only by
helical fields large enough to delay the axion oscillation,
i.e. (\ref{eq:iv}), 
unless $\abs{\alpha} $ exceeds~$8 \pi$.)
Supposing $F \tilde{F} \propto s^{4/3}$ to hold after the trapping,
and rewriting $s$ in terms of $H$ using (\ref{eq:RD}), we obtain
\begin{equation}
\left| \frac{F \tilde{F}}{f^2 H^2} \right|_{\mathrm{tr}}
=  \left( \frac{128 \pi^2}{45} \right)^{1/3}
\frac{g_{* s} (T_{\mathrm{tr}})^{4/3}}{g_{* } (T_{\mathrm{tr}})}
\frac{\abs{F \tilde{F}}_0 M_{\mathrm{Pl}}^2}{f^2 s_0^{4/3}}.
\end{equation}
Using this and setting $g_{*s}^{4/3} / g_* \sim 1$,
one finds that the backreaction is non-negligible if
\begin{equation}
 \alpha^2 \Delta N_{E \cdot B} \gtrsim
\left( \frac{f}{10^{17}\, \mathrm{GeV}}\right)^2.
\label{eq:I-vi}
\end{equation}
Note that the strength of backreaction depends on 
$\alpha^2 \Delta N_{E \cdot B}$, 
while the axion abundance is set by $\alpha \Delta N_{E \cdot B}$.

\begin{figure}[t]
\begin{center}
 \begin{minipage}{.46\linewidth}
  \begin{center}
 \includegraphics[width=\linewidth]{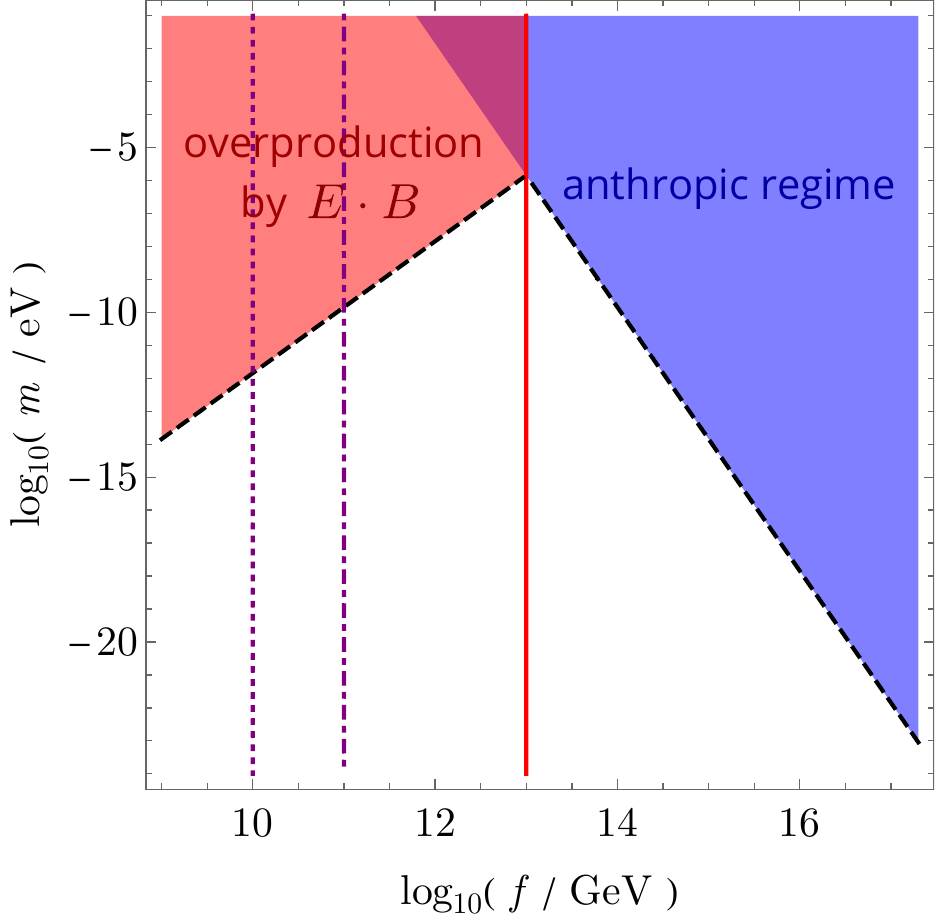}
  \end{center}
 \end{minipage} 
 \begin{minipage}{0.01\linewidth} 
  \begin{center}
  \end{center}
 \end{minipage} 
 \begin{minipage}{.46\linewidth}
  \begin{center}
 \includegraphics[width=\linewidth]{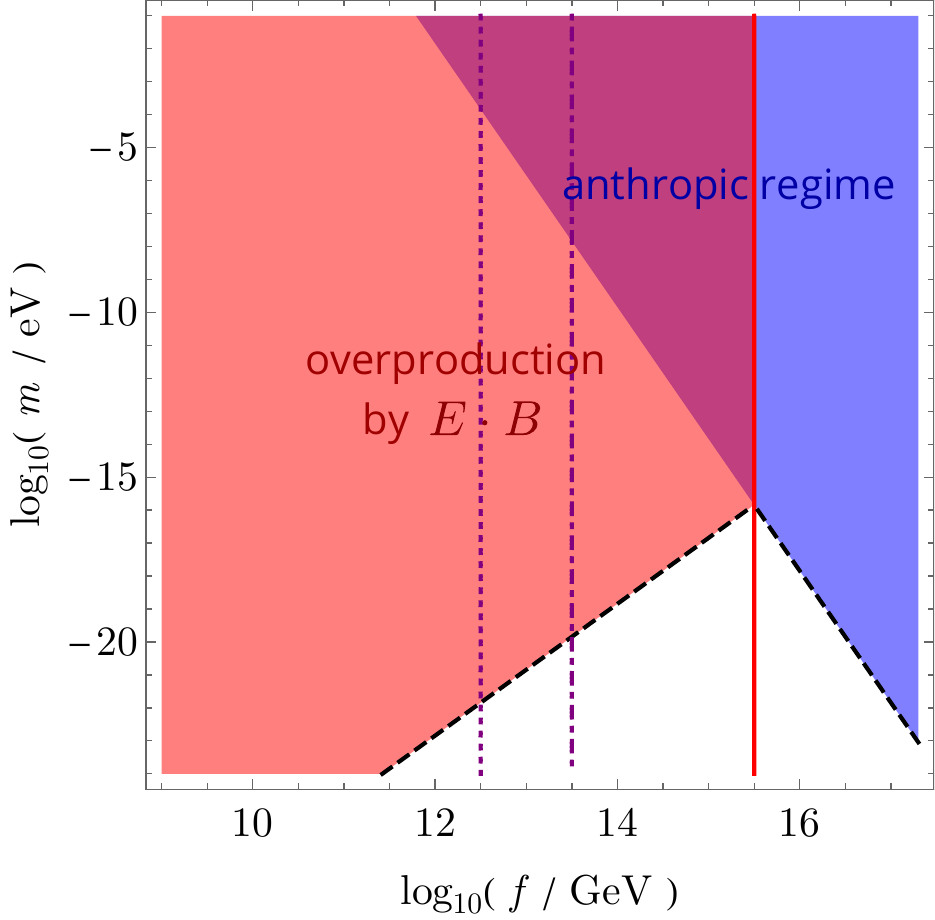}
  \end{center}
 \end{minipage} 
 \caption{Window of constant-mass axions in the presence of helical
 electromagnetic fields.
 Axes are the axion decay constant and mass, and the amplitude of the
 helical fields are taken as $\abs{\alpha} \Delta N_{E \cdot
 B} = 10^{-10}$ (left panel), $10^{-5}$ (right).
 The axion makes up the entire 
 dark matter density on the black dashed line. The helical
 fields enhance the axion abundance on the left side of the red solid
 line, and overproduces axion dark matter in the red region. 
 The blue region shows where axion dark matter is overproduced in the
 conventional vacuum misalignment scenario, unless the misalignment
 angle is tuned. 
 Backreactions from the axion to the helical fields are
 non-negligible on the left side of the purple dotted (dot-dashed)
 lines for $\abs{\alpha} = 10^{-4}$ ($10^{-2}$).}
 \label{fig:axionlike}
\end{center}
\end{figure}

In Figure~\ref{fig:axionlike}, we show the axion window in terms of $f$
and $m$. 
The left panel is the case with 
$\abs{\alpha} \Delta N_{E \cdot B} = 10^{-10}$,
and the right is
$\abs{\alpha} \Delta N_{E \cdot B} = 10^{-5}$.
The lower edge of the blue region shows the parameter combinations
in the conventional misalignment scenario (cf. (\ref{eq:alp-vm}))
for which an axion makes up the entire dark matter abundance, 
i.e. $\Omega_\theta h^2 \approx 0.1$,
with an initial angle $\abs{\theta_m} \approx 1$.
Inside the blue region, the axion gives too much dark matter in the
universe, unless the initial angle takes fine-tuned values of
$\abs{\theta_m} \ll 1$ (the ``anthropic window'').
The condition of (\ref{eq:iv}) is satisfied in the left side of the
red solid line, where the onset of the axion oscillation is delayed
by the helical electromagnetic fields and thus the relic abundance is
given by (\ref{eq:I-ii}). 
Here, $\Omega_\theta h^2 \approx 0.1$ is realized on the lower edge of
the red region, 
while inside the red region axion dark matter is overproduced.
The parameter window affected by the helical fields expands to larger
$f$ with increasing $\abs{\alpha} \Delta N_{E \cdot B}$. 
Putting together the regions affected/unaffected by the helical
fields, the black dashed line shows where $\Omega_\theta h^2
\approx 0.1$ without a fine-tuned initial condition.
However we should remark that the backreaction from the axion to the
helical fields becomes non-negligible, i.e. (\ref{eq:I-vi}), 
on the left side of the  
purple dotted line for $\abs{\alpha} = 10^{-4}$,
and the purple dot-dashed line for $\abs{\alpha} = 10^{-2}$.
(For these values of $\abs{\alpha}$, the choices of $\abs{\alpha} \Delta
N_{E \cdot B}$ in the plots imply $\Delta N_{E \cdot B} \leq 10^{-1}$, 
being compatible with the observational constraint~(\ref{eq:NEB}) on extra
radiation.
The rate of decay of the axion into hidden photons via $(\alpha / 8 \pi)
\theta F \tilde{F}$ is also small such that the axion lifetime is longer
than the age of the universe in the entire parameter range shown in the
plots.)
As the gauge coupling~$\abs{\alpha}$ becomes larger, 
the strong-backreaction region spreads towards larger~$f$.
Here the axion is expected to induce a UV cascading of the
helical fields and thus the calculation (\ref{eq:I-ii}) of the abundance
may become invalid.\footnote{The calculation may also break down when
$\Delta N_{E \cdot B}$ is so large that the hidden photon and/or the
axion dominate over the SM radiation already in the early
universe, hence violating our assumption of radiation domination
at $H = m$.}
We also note that
in the plots we have used the
order-of-magnitude estimates (\ref{eq:I-ii}) and (\ref{eq:alp-vm}), 
which ignore the possibility of the initial misalignment angle and
$\abs{\alpha} F\tilde{F}$ conspiring to give 
anharmonic enhancements of the axion abundance,
as depicted by the oscillation peaks in the right panel of
Figure~\ref{fig:num-theta}. 
This effect can further transform the axion window.

Let us also comment on the implication for
ultralight axion dark matter~\cite{Hu:2000ke,Hui:2016ltb} with
$m \lesssim 10^{-21}\, \mathrm{eV}$, which is constrained by studies of the
Lyman-$\alpha$
forest~\cite{Irsic:2017yje,Armengaud:2017nkf,Kobayashi:2017jcf}
and galaxy rotation curves~\cite{Bar:2018acw}.
For axions in this mass regime to make up most of the dark matter,
the conventional misalignment scenario
requires $f \gtrsim 10^{17}\, \mathrm{GeV}$.
However with an $F\tilde{F}$ background, axions with much smaller decay
constants can also account for ultralight dark matter and leave distinct
signatures in the small-scale structures of the universe.

\subsection{QCD Axion}
\label{sec:QCDaxion}

We now study the impact of an $F\tilde{F}$ background on the QCD
axion~\cite{Peccei:1977hh,Weinberg:1977ma,Wilczek:1977pj},
whose mass depends on the cosmic temperature approximately as
\begin{equation}
  m (T) \simeq
 \begin{dcases}
 \lambda \,  m_{0} \left( \frac{\Lambda_{\mathrm{QCD}} }{T} \right)^p
  & \mathrm{for}\, \, \,  T \gg \Lambda_{\mathrm{QCD}}, \\
 m_{0}
  & \mathrm{for}\, \, \,  T \ll \Lambda_{\mathrm{QCD}}.
 \end{dcases}
\label{eq:qcd_mass}
\end{equation}
Here $\Lambda_{\mathrm{QCD}} \approx 200\, \mathrm{MeV}$, 
$\lambda \approx 0.1$, $p \approx 4$, and the zero-temperature mass is
given by \cite{diCortona:2015ldu,Borsanyi:2016ksw}
\begin{equation}
 m_0 \approx 6 \times 10^{-6} \, \mathrm{eV} 
\left(\frac{10^{12} \, \mathrm{GeV}}{f}\right).
\label{eq:zero_mass}
\end{equation}

Let us again suppose the helical fields to be composed of hidden photons
with a redshifting behavior $F\tilde{F} \propto a^{-4}$,
and that the universe becomes radiation-dominated by the time 
when $H = m$. 
The calculations can be carried out similarly to the previous section,
except for that now the axion mass also varies in time.
We focus on axions with decay constants of
$ f \lesssim 10^{17}\, \mathrm{GeV}$, for which
the Hubble rate becomes equal to the axion mass at temperatures
$T_m \gtrsim \Lambda_{\mathrm{QCD}}$, giving
$m(T_m) \simeq \lambda m_0 (\Lambda_{\mathrm{QCD}} / T_m)^p$.

We begin by considering $F\tilde{F}$ whose amplitude is so
large that the onset of the axion oscillation is delayed to times when
the cosmic temperature has dropped below the QCD scale, 
i.e. $T_{\mathrm{tr}} \lesssim \Lambda_{\mathrm{QCD}} \lesssim T_m$.
For this case the mass is already a constant when the oscillation begins, 
i.e., $m(T_{\mathrm{tr}}) \simeq m_0$,
so the relic abundance for constant-mass axions (\ref{eq:I-ii}) 
applies by simply replacing $m \to m_0$.
However the lower bound on $F \tilde{F}$ is now stronger,
as we are imposing 
$T_{\mathrm{tr}} \lesssim \Lambda_{\mathrm{QCD}}$.
We slightly modify this condition to 
$\lambda (\Lambda_{\mathrm{QCD}} / T_{\mathrm{tr}})^p > 1$
so that 
$m(T_{\mathrm{tr}}) \simeq m_0$ under the expression~(\ref{eq:qcd_mass});
then by solving for $T_{\mathrm{tr}}$ as discussed around~(\ref{eq:4.9}),
one finds the bound
\begin{equation}
 \left| \frac{\alpha }{8 \pi } (F\tilde{F})_0  \right|
> \frac{1}{2} \left(\frac{45}{2 \pi^2}\right)^{5/6}
\frac{g_*(T_{\mathrm{tr}})^{1/2}}{g_{*s}(T_{\mathrm{tr}})^{4/3}}
\frac{m_0 f^2 s_0^{4/3} }{\lambda^{2/p} \Lambda_{\mathrm{QCD}}^2 M_{\mathrm{Pl}} }.
\label{eq:II-a-ex}
\end{equation}
Also noting that the zero-temperature mass of the QCD axion is a
function of the decay constant, cf.~(\ref{eq:zero_mass}),
and plugging in numbers,
the abundance is obtained as
\begin{equation}
 \Omega_\theta h^2 \sim 10^{-1}
\left( \frac{\abs{\alpha } \Delta N_{E \cdot B}}{10^{-11}}
\right)^{3/2}
\left( \frac{f}{10^{12}\, \mathrm{GeV}} \right)^{-3/2}
\quad
\textrm{for}
\, \, \, 
\abs{\alpha} \Delta N_{E \cdot B} \gtrsim
 10^{-6} \, 
\left( \frac{f}{10^{12}\, \mathrm{GeV}} \right).
\label{eq:II-ii}
\end{equation}

Even if $F\tilde{F}$ is not as large as shown in (\ref{eq:II-ii}), it
still impacts the QCD axion abundance as long as it is large enough to 
delay the onset of the oscillation to times when $H < m$, 
namely, if $\Lambda_{\mathrm{QCD}} \lesssim T_{\mathrm{tr}} < T_m$. 
In this case the mass continues to vary in time after the axion trapping.
The lower bound for $F\tilde{F}$ arises from (\ref{eq:iv}), 
which turns out to take exactly the same form as (\ref{eq:I-iv-ex}). 
One can solve for $T_{\mathrm{tr}}$ using 
$m(T_{\mathrm{tr}}) \simeq \lambda m_0 (\Lambda_{\mathrm{QCD}} /
T_{\mathrm{tr}})^p $,
and derive the relic abundance as
\begin{equation}
 \rho_{\theta 0}  = c 
\left\{
\frac{2 \pi^2}{45}
\frac{g_{*s} (T_{\mathrm{tr}})}{\lambda \Lambda_{\mathrm{QCD}}^p m_0^2
f^2  s_0}  
\right\}^{\frac{1}{p+2}}
\left\{ 2
 \left(\frac{45}{2 \pi^2}\right)^{1/6}
\frac{g_{*s} (T_{\mathrm{tr}})^{1/3}}{g_{*} (T_{\mathrm{tr}})^{1/2}}
\left| \frac{\alpha }{8 \pi }(F \tilde{F})_0 \right| 
\frac{m_0 M_{\mathrm{Pl}}}{s_0^{1/3}} 
\right\}^{\frac{p+3}{p+2}}.
\label{eq:III-ii-ex}
\end{equation}
Thus we find
\begin{equation}\label{eq:III-ii}
\begin{split}
 \Omega_\theta h^2 \sim
10^{-1}
\left( \frac{\abs{\alpha } \Delta N_{E \cdot B}}{10^{-12}}
\right)^{7/6}
&\left( \frac{f}{10^{12}\, \mathrm{GeV}} \right)^{-7/6}
\\
&\mathrm{for}
\, \, \, 
 10^{-12} \,
\left( \frac{f}{10^{12}\, \mathrm{GeV}} \right)^2
\lesssim
\abs{\alpha} \Delta N_{E \cdot B}
\lesssim
 10^{-6} \, 
\left( \frac{f}{10^{12}\, \mathrm{GeV}} \right).
\end{split}
\end{equation}
The lower bound on $\abs{\alpha} \Delta N_{E \cdot B}$ is
equivalent to that in (\ref{eq:I-ii}), however 
the reference value for~$f$ has been changed.

An even smaller $F\tilde{F}$ merely shifts the axion field
value prior to the onset of the oscillation at $H \sim m$, 
and the abundance takes the familiar expression (\ref{eq:rho_conv})
from the conventional misalignment scenario~\cite{Turner:1985si}: 
\begin{equation}
 \Omega_\theta h^2 \sim
10^{-1} \,  \theta_m^2 \, 
\left( \frac{f}{10^{12}\, \mathrm{GeV}} \right)^{7/6}
\quad
\textrm{for}
\, \, \,
\abs{\alpha} \Delta N_{E \cdot B}
\lesssim
  10^{-12} \,
\left( \frac{f}{10^{12}\, \mathrm{GeV}} \right)^2.
\label{eq:QCD-vm}
\end{equation}

In summary, the relic abundance of the QCD axion takes the forms
(\ref{eq:II-ii}), (\ref{eq:III-ii}), and (\ref{eq:QCD-vm}),
depending on the amplitude of the $F\tilde{F}$ background. 
One immediately finds that the expression~(\ref{eq:II-ii}) 
in the large $\abs{\alpha} \Delta N_{E \cdot B}$ regime
gives $\Omega_\theta h^2 \gtrsim 10^6$,
indicating an overproduction of axion dark matter.
(Note that this simply implies that the axion
density would be much larger than the measured value of the critical
density, but it does not mean that the axion would actually
``overclose'' the universe.) 
We also note that the condition for significant
backreaction~(\ref{eq:I-vi}) applies for the QCD axion in exactly the same form, as
its derivation is independent of the time evolution of the mass.

\begin{figure}[t]
 \begin{center}
 \begin{center}
 \includegraphics[width=.48\linewidth]{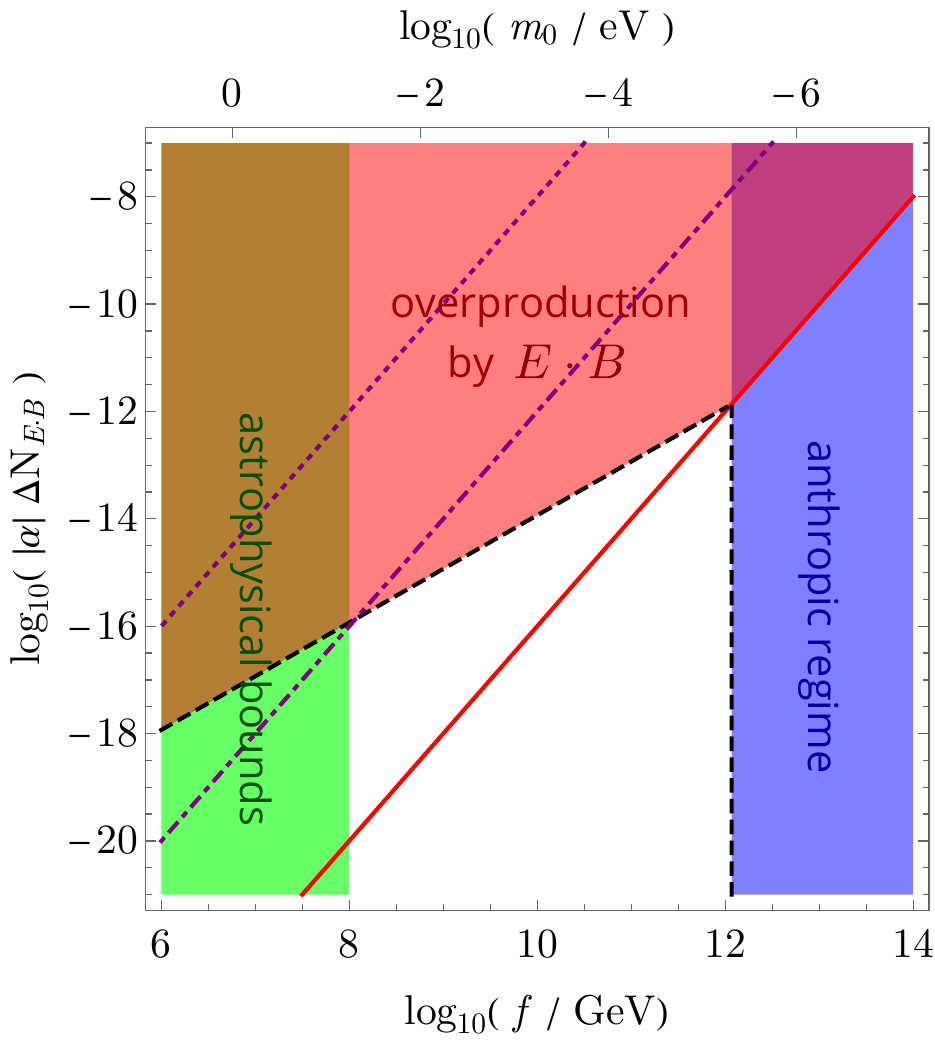}
 \end{center}
 \caption{QCD axion window in the presence of helical electromagnetic
  fields.
  Axes are the axion decay constant (bottom), zero-temperature mass
  (top), and the amplitude of the helical fields (left). 
  The axion makes up the entire dark matter density on the black dashed
  line. The helical fields enhance the axion abundance on
  the left side of the red solid line, and overproduce axion dark matter
  in the red region. 
  The blue region shows where axion dark matter is overproduced in the
  conventional vacuum misalignment scenario, unless the misalignment
  angle is tuned. 
  The green region is excluded from astrophysical observations.
  Backreactions from the axion to the helical fields are non-negligible
  on the left side of the purple dotted (dot-dashed) lines for
  $\abs{\alpha} = 10^{-6}$ ($10^{-2}$).}
 \label{fig:QCDaxion}
 \end{center}
\end{figure}

In Figure~\ref{fig:QCDaxion}, we show the axion window in terms of
$f$ and $\abs{\alpha} \Delta N_{E \cdot B}$.
On the left edge of the blue region, i.e. $f \sim 10^{12}\, \mathrm{GeV}$,
the axion makes up the entire dark matter abundance
in the conventional misalignment scenario (cf. (\ref{eq:QCD-vm})),
with an initial angle $\abs{\theta_m} \approx 1$.
Inside the blue region, axion dark matter is overproduced 
unless the initial angle is tuned to $\abs{\theta_m} \ll 1$. 
The condition (\ref{eq:iv}) is satisfied in the left side of the
red solid line, where the relic abundance directly depends on the
helical field amplitude as (\ref{eq:II-ii}) or (\ref{eq:III-ii}). 
$\Omega_\theta h^2 \approx 0.1$ is realized 
on the lower edge of the red region, where the abundance is given by 
(\ref{eq:III-ii}), 
and inside the red region axion dark matter is overproduced.
Putting together the regions affected/unaffected by the helical
fields, the black dashed line shows where $\Omega_\theta h^2
\approx 0.1$ without a fine-tuned initial condition.
The backreaction from the axion to the
helical fields become non-negligible, i.e. (\ref{eq:I-vi}), 
on the left side of the  
purple dotted line for $\abs{\alpha} = 10^{-6}$,
and the purple dot-dashed line for $\abs{\alpha} = 10^{-2}$.
(These values of $\abs{\alpha}$ give $\Delta N_{E \cdot B} \leq 10^{-1}$
and a negligibly small decay rate of the axion into hidden photons 
in the entire displayed area.)
In these backreaction regions the results (\ref{eq:II-ii}) and
(\ref{eq:III-ii}) may become invalid.
The green region is excluded by astrophysical studies setting lower
bounds of $f \gtrsim 10^8\, \mathrm{GeV}$
on the QCD axion decay constant~\cite{Zyla:2020zbs}. 
For axions marginally satisfying this
astrophysical bound, one sees from the plot that the oscillation is
delayed by helical fields as small as  
$\abs{\alpha} \Delta N_{E \cdot B} \sim 10^{-20}$.
Moreover, for $\abs{\alpha} \Delta N_{E \cdot B} \gtrsim 10^{-12}$,
the conventional QCD axion window 
$10^8\, \mathrm{GeV} \lesssim f \lesssim 10^{12}\, \mathrm{GeV}$
is completely closed.\footnote{This can also be derived from
(\ref{eq:III-ii}) and (\ref{eq:QCD-vm}) which indicate that, 
for $\theta_m^2 \sim 1$ and a fixed $\abs{\alpha} \Delta N_{E \cdot B}$, 
the axion abundance is minimized at 
$f \sim 10^{18}\, \mathrm{GeV} \, (\abs{\alpha} \Delta N_{E \cdot
B})^{1/2}$ 
as $(\Omega_\theta h^2)_{\mathrm{min}} \sim 10^6 (\abs{\alpha} \Delta
N_{E \cdot B})^{7/12}$.} 
In the conventional misalignment scenario,
for axions with $f < 10^{12}\, \mathrm{GeV}$ to make up
the entire dark matter abundance,
one would need to invoke a tuned initial condition 
of $\abs{\theta}_m \approx \pi$ so that the abundance receives
anharmonic enhancements,
although this would also enhance isocurvature perturbations and thus is
constrained by CMB measurements~\cite{Kobayashi:2013nva}. 
We find that helical electromagnetic fields can allow 
such low-$f$ axions to be abundantly produced without invoking
anharmonic effects.

\section{Conclusions and Discussion}
\label{sec:conc}

Our main message is that primordial electromagnetic fields can strongly
impact the cosmic evolution of axions, and vice versa. 
In this paper, we have shown that helical electromagnetic fields
coherently excited in 
the early universe induce a field velocity to the coupled axion. 
If the amplitude of $E \cdot B$ is large enough such that the
induced kinetic energy of the axion is larger than the height of the axion
potential at the time when $H \sim m$, 
then the axion continues to roll in one direction and 
the onset of the axion oscillation is delayed, 
leading to an enhancement of the final axion abundance. 
For a QCD axion coupled to a hidden U(1) gauge field, 
the abundance was derived as
(\ref{eq:II-ii}), (\ref{eq:III-ii}), and (\ref{eq:QCD-vm}),
depending on the helical field amplitude.
In particular, we found that the 
conventional window 
$10^8\, \mathrm{GeV} \lesssim f \lesssim 10^{12}\, \mathrm{GeV}$
becomes completely closed
due to overproduction of axion dark matter,
with helical fields as small as
$\abs{\alpha} \Delta N_{E \cdot B} \gtrsim 10^{-12}$,
where $\alpha$ is the gauge coupling and $\Delta N_{E \cdot B}$ 
parametrizes the amplitude of $ E \cdot B$ in terms of the effective
extra relativistic degrees of freedom. 
Comparing with the current observational constraint on extra radiation
$\Delta N_{E \cdot B} \lesssim 10^{-1}$,
there is a wide range of parameters for which hidden helical
fields trigger an overproduction of the QCD axion. 
For an axion-like particle with a constant mass, its abundance is
given by (\ref{eq:I-ii}) and (\ref{eq:alp-vm}).
The window of axion-like particles was also found to be significantly
modified by rather weak helical fields.
One implication of the results is that, for axions in the ultralight
mass range $m \lesssim 10^{-21}\, \mathrm{eV}$ to make up the entire
dark matter and leave distinct signatures in the small-scale structures
of the 
universe, 
the required value of the decay constant in the presence of an $E \cdot
B$ becomes $f \ll 10^{17}\, \mathrm{GeV}$, being much smaller than in
the conventional misalignment scenario.
Our results can further be used to put constraints on early universe
models that give rise to helical electromagnetic fields, from the
requirement that axion dark matter not be overproduced.

In order to explicitly compute the axion abundance, we focused on a
case where the axion couples to a hidden U(1) field; this field was
considered to make up helical electromagnetic fields that are
homogeneous over the entire 
observable universe, and whose dot product redshifts as 
$E \cdot B \propto a^{-4}$.
We also supposed the spontaneous breaking of the global U(1) symmetry
to happen before inflation.
Let us comment on each of these assumptions and discuss directions for
further 
study. 
\begin{itemize}
 \item {\it SM/hidden, Abelian/non-Abelian gauge field.} 
       We mainly considered a hidden U(1) gauge field that is
       decoupled from the SM, such that the hidden electric field survives
       the reheating epoch. However we remark that the SM electromagnetic
       field can also enhance the axion abundance, if the reheating
       temperature is sufficiently low and the conductivity of the
       universe stays tiny\footnote{The misalignment scenario with late
       reheating is discussed in
       \cite{Lazarides:1987zf,Banks:1996ea,Giudice:2000ex}. We should also 
       note that the evolution of the conductivity depends
       not only on the reheating temperature, but also on the details of
       the reheating process.} until the time when $H \sim m$.
       It would also be interesting to apply our discussions to
       non-Abelian gauge fields and study effects from, for instance,
       color electromagnetic fields. 
 \item {\it Homogeneous/inhomogeneous electromagnetic fields.}
       The actual spectrum of primordial electromagnetic fields
       depends on the excitation mechanism of the gauge fields.
       We have considered the Fourier modes of the electromagnetic fields to be
       concentrated at wavelengths larger than the present Hubble
       radius, and studied uniform effects on the axion in the entire
       visible universe. 
       If instead the fields have
       small-wavelength components, then the axion is forced to move
       differently in each patch of the universe. This
       gives rise to isocurvature perturbations of the axion density,
       and even the formation of axionic domain walls if the
       helical field amplitude is sufficiently large.       
       Inhomogeneous electromagnetic fields also imply that the
       (dark) radiation density itself fluctuates, which may be
       constrained from observations.
 \item {\it Redshifting of helical electromagnetic fields.}
       We discussed in Appendix~\ref{app:evolution} that $E \cdot
       B$ may exhibit a redshifting behavior other than the
       radiation-like $\propto
       a^{-4}$, depending on how it is originally produced. 
       In such cases, the dependence of the axion abundance on $\Delta
       N_{E \cdot B}$ would be modified from our results. 
 \item {\it Spontaneous symmetry breaking before/after inflation.}
       We focused on cases where the global U(1) breaks before
       inflation, and studied how the vacuum misalignment scenario is
       affected by the presence of helical electromagnetic fields. 
       Here we note that the conventional misalignment scenario assumes
       that during inflation the Hubble rate is larger than the axion
       mass; otherwise the axion would be stabilized at the potential
       minimum and thus the axion particles cannot be produced.
       However, even with a low-scale inflation whose  Hubble rate is
       smaller than the axion mass, 
       helical electromagnetic fields can kick the axion out of the
       vacuum and produce axion dark matter.
       (See also \cite{Kobayashi:2019eyg,Kobayashi:2020ryx} for a
       closely related idea.)
       On the other hand, a symmetry breaking after inflation gives rise
       to cosmic strings and domain walls.
       Helical fields should also have an effect on this scenario, 
       especially on the formation of the walls due to the periodic axion
       potential. 
 \item {\it Generation of helical electromagnetic fields.}
       The primary goal of the present work is to illustrate the impact of
       helical electromagnetic fields on the axion evolution, and hence
       we did not specify how the helical fields have been generated.
       Clearly an important direction for further work is investigating
       the generation mechanisms, which is also required for
       understanding the redshifting behavior of the helical fields in
       the later universe. 
       For a recent study of a generation model for large-scale helical
       fields, see e.g. \cite{Fujita:2019pmi}.
 \item {\it Axion-induced UV cascade of helical electromagnetic fields.}
       Our simple estimates suggest that an axion induces helical
       electromagnetic fields to undergo a UV cascade, 
       when the condition~(\ref{eq:17.2}) is satisfied.
       If true, this would provide an absolute upper bound on the
       amplitude of large-scale helical electromagnetic fields 
       in our universe, in terms of the axion coupling.
       It is also important to understand the axion abundance in
       this cascading regime.
 \item {\it Further imprints.}
       In addition to affecting the axion field dynamics, primordial
       (hidden) electromagnetic fields can induce conversions
       between axions and (hidden) photons~\cite{Sikivie:1983ip,Kamada:2017cpk}.
       The helical electromagnetic background can also leave
       parity-violating signatures in cosmological observations
       (see e.g. \cite{Pogosian:2001np} for signals from helical SM
       magnetic fields).
       It will be interesting to systematically analyze the range of
       signals arising in a universe with axions and primordial
       electromagnetic fields. 
\end{itemize}
The effects listed here can provide us with further opportunities to probe
the interplay between axions and primordial electromagnetic fields. 
We leave a detailed study of them to future work.

\section*{Acknowledgments}

We thank Martin Sloth and Lorenzo Ubaldi for useful comments on a draft
of this paper, and Kiyotomo Ichiki and Hiroyuki Tashiro for 
helpful discussions.
RKJ would like to acknowledge financial support from the new faculty seed start-up grant of IISc, the Core Research Grant CRG/2018/002200 from the Science and Engineering Research Board, Department of Science and Technology, Government of India and the Infosys Foundation, Bangalore.


\appendix

\section{Early Cosmological Evolution of Helical Electromagnetic Fields}
\label{app:evolution}

We study the cosmological evolution of helical electromagnetic fields
that have been produced in the early universe, and show that they do not
necessarily redshift in a radiation-like manner in a vacuum as one would
naively expect. 
The discussion here closely follows the work~\cite{Kobayashi:2019uqs}.

\subsection{Model with a Time-Dependent and Parity-Violating Background} 

In this appendix we assume the helical electromagnetic fields to 
arise from a background that spontaneously breaks time
diffeomorphisms as well as parity. 
As a simple toy example, we study the following U(1) gauge field theory,
\begin{equation}
 S = \frac{1}{4} \int d^4 x \sqrt{-g} 
\left[ 
-I(\tau)^2  F_{\mu \nu}  F^{\mu \nu}
+ J(\tau) F_{\mu \nu} \tilde{F}^{\mu \nu}
 \right].
\label{eq:A.1}
\end{equation}
The dual field strength is defined as in (\ref{eq:Ftilde}),
and we fix the metric to a flat FRW,
$ds^2 = a(\tau)^2 ( -d\tau^2 + d\bd{x}^2 )$.
The explicit time dependence of the coefficients $I$ and $J$ 
are understood to arise from
couplings to other coherent degrees of freedom, such
as a time-evolving scalar 
(see e.g. \cite{Turner:1987bw,Ratra:1991bn,Garretson:1992vt,Anber:2006xt,Durrer:2010mq,Byrnes:2011aa,Adshead:2016iae,Ferreira:2013sqa,Ferreira:2014hma}
for explicit models).
These time-dependent coefficients violate the Weyl invariance of the
Yang--Mills action and enables the production of electromagnetic fields
in a Weyl-flat spacetime. The term $J(\tau) F \tilde{F}$ further violates
parity and thus allows the fields to be helical.\footnote{A general
gauge field theory with spontaneously broken time 
diffeomorphisms can further include terms such as
$\mathcal{K} (\tau) \tensor{F}{^{0}_{\mu}} F^{0 \mu } $,
which yields a variable speed of light~\cite{Green:2015fss}.
However the model of (\ref{eq:A.1}) is good enough for our purpose of
studying the redshifting behaviors of electromagnetic fields after
they are produced.} 
$I(\tau)^2$ is assumed to be positive definite.

The spatial components of the gauge field can be decomposed into
irrotational and incompressible parts, 
\begin{equation}
 A_\mu = (A_0, \partial_i S + V_i)
 \quad  \mathrm{with} \quad
  \partial_i V_i = 0,
\end{equation}
where a sum over
repeated spatial indices is implied irrespective of their positions. 
Noting that $A_0$ is a Lagrange multiplier whose constraint equation
under proper boundary conditions gives $A_0 = S'$
(a prime denotes a derivative with respect to the conformal time~$\tau$),
one can eliminate both $A_0$ and $S$
from the action to yield, up to surface terms,
\begin{equation}
 S = \frac{1}{2} \int d \tau d^3 x
\left[
I^2  ( V_i' \, V_i' - \partial_i V_j  \, \partial_i V_j)
- J'  \epsilon_{0ijl}  V_i \partial_j V_l
\right].
\end{equation}
Here we have written the totally antisymmetric pseudotensor as 
$\eta_{\mu \nu \rho \sigma} = - \sqrt{-g} \epsilon_{\mu \nu \rho
\sigma}$
using a Levi--Civita symbol with $\abs{\epsilon_{0123}} = 1$
and $\epsilon_{0123} = \epsilon^{0123}$.

We introduce two orthonormal polarization vectors
$e_i^{1}(\boldsymbol{k})$ and $e_i^{2}(\boldsymbol{k})$ that satisfy
\begin{equation}
 e_i^{p} (\bd{k}) \,  k_i = 0, 
\quad
 e_i^{p} (\bd{k}) \,  e_i^{q} (\bd{k})  =
 \delta^{pq},
\quad
\mathrm{for} \, \, \, 
p, q = \left\{ 1, 2 \right\},
\end{equation}
and their combinations,
\begin{equation}
 e_i^{\pm} (\bd{k}) = \frac{1}{\sqrt{2}}
\left( e_i^{1}(\bd{k}) \pm i e_i^{2} (\bd{k})\right).
\end{equation}
It can be checked that these vectors satisfy
\begin{equation}
e_i^{+} (\bd{k}) \, e_j^{-} (\bd{k})
+ e_i^{-} (\bd{k}) \, e_j^{+} (\bd{k})
 = \delta_{ij} - \frac{k_i k_j}{k^2},
\quad
 \epsilon_{0ijl} k_j e_l^{\pm} (\bd{k})
= \mp i k e_i^{\pm} (\bd{k}).
\end{equation}
Here $k \equiv \sqrt{k_i k_i}$, 
and we have chosen the directions of the polarization vectors such
that the signs in the right hand side of the second equation are as shown. 
We further require the polarization vectors to satisfy 
\begin{equation}
 e_i^{1} (\bd{k})^* = e_i^{1} (\bd{k}),
\quad
 e_i^{2} (\bd{k})^* = e_i^{2} (\bd{k}),
\quad
  e_i^{1} (-\bd{k}) = e_i^{1} (\bd{k}),
\quad
  e_i^{2} (-\bd{k}) = - e_i^{2} (\bd{k}),
\end{equation}
so that 
\begin{equation}
 e_i^{\pm} (\bd{k})^* = e_i^{\mp} (\bd{k}),
\quad
 e_i^{\pm} (-\bd{k}) = e_i^{\mp} (\bd{k}).
\end{equation}

\subsection{Canonical Quantization}

We now quantize the gauge field by promoting $V_i$ to an operator,
\begin{equation}
\begin{split}
 V_i(\tau, \boldsymbol{x}) = 
 \int \frac{d^3 k}{(2 \pi)^3} 
\Bigl[
& e^{+}_i (\bd{k})
\left\{
e^{i \bd{k \cdot x}}  a_{\bd{k}}
u^{+}_{\bd{k}} (\tau) + 
e^{-i \bd{k \cdot x}} b_{\bd{k}}^{\dagger}
u^{-}_{\bd{k}} (\tau)^*  
\right\} \\
+ & e^{-}_i (\bd{k})
\left\{
e^{i \bd{k \cdot x}}  b_{\bd{k}}
u^{-}_{\bd{k}} (\tau) + 
e^{-i \bd{k \cdot x}} a_{\bd{k}}^{\dagger}
u^{+}_{\bd{k}} (\tau)^*
\right\}
\Bigr],
\end{split}
\end{equation}
where we take the mode functions $u_{\bd{k}}^{\pm} (\tau)$ to 
obey the equations of motion,
\begin{equation}
u_{\bd{k}}^{\pm \prime \prime} + 2 \frac{I'}{I} u_{\bd{k}}^{\pm\prime} + 
\left( k^2 \pm \frac{J'}{I^2}k \right) u_{\bd{k}}^{\pm} = 0,
\label{eq:u-EoM}
\end{equation}
and the time-independent annihilation and creation operators to 
have the commutation relations,
\begin{equation}
\begin{split}
&[ a_{\bd{k}},\,  a_{\bd{q}}^{\dagger} ] = 
 [ b_{\bd{k}},\,  b_{\bd{q}}^{\dagger} ] = 
(2  \pi)^3 \, \delta^{(3)}  (\bd{k} - \bd{q}), \\
&[ a_{\bd{k}},\,  a_{\bd{q}} ] =
 [ b_{\bd{k}},\,  b_{\bd{q}} ] =
 [ a_{\bd{k}},\,  b_{\bd{q}} ] =
 [ a_{\bd{k}},\,  b_{\bd{q}}^{\dagger} ] =
 \cdots 
 = 0.
\end{split}
\end{equation}
We also take $V_i$ and its conjugate
momentum obtained from the action $S = \int d\tau d^3x \mathcal{L}$ as
$ \Pi_i = \partial \mathcal{L} / \partial V_i'  = I^2 V_i'$,
to obey the equal-time commutation relations
\begin{equation}
 \begin{split}
& \left[ V_i(\tau, \bd{x}),\,  \Pi_j (\tau, \bd{y}) \right] 
  = i \int \frac{d^3 k}{(2\pi)^3} \, 
 e^{i\bd{k\cdot}  (\bd{x - y})}
\left(
\delta_{ij} - \frac{k_i k_j}{k^2}
\right),
\\
& \left[ V_i(\tau, \bd{x}),\,  V_j (\tau, \bd{y}) \right] = 
 \left[ \Pi_i(\tau, \bd{x}),\,  \Pi_j (\tau, \bd{y}) \right]
 = 0.
 \end{split}
\end{equation}
The two set of commutation relations are consistent with each other when 
the mode functions are independent of the direction of~$\boldsymbol{k}$,
i.e., $ u_{\bd{k}}^{\pm} = u_k^{\pm}$,
and obey the normalization condition
\begin{equation}
 I^2 \left(
u_k^{\pm} u^{\pm \prime *}_k - u_k^{\pm *} u^{\pm \prime}_k
\right) = i.
\label{eq:norm}
\end{equation}

Then, defining the vacuum state by 
$a_{\bd{k}} |0 \rangle = b_{\bd{k}} |0 \rangle = 0$
for $^{\forall}  \boldsymbol{k}$,
we can compute the correlation functions of the electromagnetic
fields~(\ref{eq:EandB}),
\begin{equation}
  \langle 0| X_\mu (\tau, \boldsymbol{x}) Y^\mu (\tau, \boldsymbol{y}) |0
 \rangle =
 \int \frac{d^3 k}{4 \pi k^3}  e^{i\boldsymbol{k\cdot}  (\boldsymbol{x
- y})} \mathcal{P}_{X Y} (\tau, k),
\quad
\mathrm{where}
\, \, \, 
X, Y = \left\{ E, B  \right\}.
\end{equation}
The power spectrum for each combination of the fields are given in
terms of the mode functions as
\begin{equation}
\mathcal{P}_{EE} = \frac{k^3}{2 \pi^2 a^4} 
\sum_{p=\pm} | u^{p \, \prime}_k |^2,
\quad
 \mathcal{P}_{BB} = \frac{k^5}{2 \pi^2 a^4} 
\sum_{p=\pm} | u_k^{p} |^2,
\quad
\mathcal{P}_{EB} = \frac{k^4}{4 \pi^2 a^4} 
\left( \abs{u_k^-}^2 - \abs{u_k^+}^2 \right)'.
\label{eq:P_XX}
\end{equation}
We thus clearly see that a correlation between the electric and
magnetic fields $\mathcal{P}_{EB}$ reflects an asymmetry between the two
helicity modes.
If the two modes start from the same initial condition (vacuum),
then the difference can arise only from the difference in their
time evolutions, which is sourced by the time dependence of $J$,
as seen in the equation of motion~(\ref{eq:u-EoM}).

\subsection{Redshifting of Electromagnetic Fields within Maxwell Theory}

Let us assume that after the helical electromagnetic fields have been 
produced by the time-varying $I$ and $J$, the coefficients approach
$I^2 = 1$ and $J' = 0$, recovering the standard Maxwell theory.
Then the mode functions reduce to a sum of plane waves,
\begin{equation}
 u_k^{\pm} = \frac{1}{\sqrt{2 k}}
  \left(  \alpha_k^{\pm} e^{-i k \tau }
  + \beta_k^{\pm} e^{i k \tau }\right),
\end{equation}
where $\alpha_k^{\pm}$ and $\beta_k^{\pm}$ are time-independent complex
numbers satisfying
$ \abs{\alpha_k^{\pm}}^2 - \abs{\beta_k^{\pm}}^2 = 1$
as required by the normalization condition~(\ref{eq:norm}).
Note that $\abs{\beta_k^{\pm}}^2$ corresponds to the number of photons
with polarization~$\pm$, per six-dimensional comoving phase volume.
Thus we have
\begin{equation}\label{eq:A.17}
\begin{split}
& \abs{u_k^{\pm}}^2 = \frac{1}{2k}
 \left[
 1 + 2 \, \abs{\beta_k^{\pm}}^2
  + 2 \, \abs{\beta_k^{\pm}} \sqrt{1 + \abs{\beta_k^{\pm}}^2}
  \, \cos \left\{ \arg (\alpha_k^{\pm} \, \beta_k^{\pm*}) - 2 k \tau
  \right\} 
\right],
\\
& \abs{u_k^{\pm \prime}}^2 = \frac{k}{2} 
 \left[
 1 + 2 \, \abs{\beta_k^{\pm}}^2
  - 2 \, \abs{\beta_k^{\pm}} \sqrt{1 + \abs{\beta_k^{\pm}}^2}
  \, \cos \left\{ \arg (\alpha_k^{\pm} \, \beta_k^{\pm*}) - 2 k \tau
  \right\} 
\right],
\\
& (\abs{u_k^{\pm}}^2)' = 
 2 \, \abs{\beta_k^{\pm}} \sqrt{1 + \abs{\beta_k^{\pm}}^2}
  \, \sin \left\{ \arg (\alpha_k^{\pm} \, \beta_k^{\pm*}) - 2 k \tau
  \right\} ,
\end{split}
\end{equation}
with which the power spectra of the electromagnetic
fields~(\ref{eq:P_XX}) can be evaluated.\footnote{Using (\ref{eq:A.17})
one also obtains
$\langle \rho_A \rangle 
= \langle E_{\mu} E^{\mu }  +  B_{\mu} B^{\mu }  \rangle/2
> \abs{ \langle E_{\mu} B^{\mu }  \rangle} $
by evaluating the correlation functions at a single spacetime point.
This inequality is slightly different from (\ref{eq:12.7}) due to the
zero-point fluctuations, 
however we remark that this computation is too naive and a more rigorous
treatment in the coincidence limit~\cite{birrell_davies_1982} would be
necessary for discussing this inequality relation at the quantum level.}
The explicit values of the constants $\abs{\beta_k^{\pm}}$
and $ \arg (\alpha_k^{\pm} \, \beta_k^{\pm*})$ depend on the gauge
field-producing model described by $I(\tau)$ and $J(\tau)$. 
Below we present two brief case studies where the
electromagnetic correlators redshift in radiation-like and
non-radiation-like manners, respectively.

\subsubsection*{Radiation-like Redshifting}

Suppose that the $+$~mode has been exclusively produced so that
$\abs{\beta_k^+}^2 \gg \abs{\beta_k^-}^2, 1$.
Further assuming that the phase
$\{ \arg (\alpha_k^{+} \, \beta_k^{+ *}) - 2 k \tau \} $ is not
close to $n  \pi $ ($n = 0, \pm 1, \pm 2, \cdots$)
such that the leading terms
in the expressions of (\ref{eq:A.17}) do not cancel each other or vanish,
then we see that the magnitudes of the three types of correlators are all of
the same order,  
\begin{equation}
 \mathcal{P}_{EE} \sim \mathcal{P}_{BB} \sim \abs{\mathcal{P}_{EB}}
  \sim \frac{k^4}{a^4} \abs{\beta_k^+}^2.
\end{equation}
In this case, all power spectra exhibit a radiation-like redshifting of 
$\mathcal{P}_{XY} \propto a^{-4}$.

\subsubsection*{Non-Radiation-like Redshifting}

For the second case, we expand the phase around $\pi$ as
\begin{equation}
 \arg (\alpha_k^{\pm} \, \beta_k^{\pm *}) - 2 k \tau 
 \equiv \pi + \theta_k^{\pm},
\end{equation}
and suppose that this phase parameter and photon density
satisfy the following relations,
\begin{equation}
 \frac{1}{\abs{\beta_k^{\pm}}^2} \ll \abs{\theta_k^{\pm}}  \ll 1,
\qquad
 \abs{\theta_k^-}^n \abs{\beta_k^-}^2 \ll
 \abs{\theta_k^+}^n \abs{\beta_k^+}^2 
\quad
\mathrm{for} \, \, \, 
n = 0,1,2.
\end{equation}
It then follows that the power spectra are approximated by  
\begin{equation}
 \mathcal{P}_{EE} \simeq \frac{k^4}{\pi^2 a^4 } 
 \abs{\beta_k^{+}}^2,
\quad
 \mathcal{P}_{BB} \simeq \frac{k^4}{4 \pi^2 a^4 } 
 (\theta_k^{+})^2 \,  \abs{\beta_k^{+}}^2,
\quad
 \mathcal{P}_{EB} \simeq \frac{k^4}{2 \pi^2 a^4 }
\theta_k^{+} \abs{\beta_k^+}^2,
\end{equation}
which entails a hierarchy 
$\mathcal{P}_{EE} \gg \abs{\mathcal{P}_{EB}} \gg \mathcal{P}_{BB}$.
(The case with an inverted hierarchy 
$\mathcal{P}_{EE} \ll \abs{\mathcal{P}_{EB}} \ll \mathcal{P}_{BB}$ can
be realized under
$\arg (\alpha_k^{\pm} \, \beta_k^{\pm *}) - 2 k \tau \approx 0$.)

Now, considering that the conformal time in an FRW universe typically
scales as $\tau \sim  (a H)^{-1} + \mathrm{const.}$
with $H = a'/a^2$ being the Hubble rate,
let us suppose the phase parameter to evolve in time as
$\theta_k^+ \propto (a H)^{-1}$.
Then it follows that the power spectra redshift as
\begin{equation}
\mathcal{P}_{EE} \propto \frac{1}{a^{4}},
\quad
\mathcal{P}_{BB} \propto \frac{1}{a^6 H^2},
\quad
\mathcal{P}_{EB} \propto \frac{1}{a^5 H}.
\end{equation}
Thus in a decelerating universe, 
the $BB$ and $EB$ power redshift slower than $a^{-4}$.
These scaling behaviors can also be understood from Faraday's law of
induction. 
For further discussions on this point, and also for explicit models that
produce electromagnetic fields with such non-radiation-like scalings,
the reader is referred to~\cite{Kobayashi:2019uqs}.

\bibliographystyle{JHEP}
\bibliography{axion}

\end{document}